\documentclass[preprint,prx,aps,amsmath,amssymb,floatfix,superscriptaddress,longbibliography]{revtex4-2}

\usepackage{mathtools,amsmath}
\usepackage{dsfont}
\usepackage{graphicx} 
\usepackage{siunitx}
\usepackage{bbold}
\usepackage{soul}
\usepackage{bm}
\usepackage[normalem]{ulem}
\usepackage{times}
\usepackage{enumerate}  
\usepackage{float}
\usepackage{amssymb}
\usepackage{xcolor} 
\usepackage{tikz}

\renewcommand{\Re}{\mathop{\text{Re}}\nolimits}
\renewcommand{\Im}{\mathop{\text{Im}}\nolimits}

\newcommand{\ket}[1]{|#1\rangle}

\newcommand{\newtext}[1]{#1}
\let\corr=\newtext

\begin{document}

%\articletype{Paper}

\title{Quantum Walk on a Line with Absorbing Boundaries}

\author{Ammara Ammara}
\affiliation{Dipartimento Interuniversitario di Fisica, Universit\`a di Bari, I-70126 Bari, Italy}
\affiliation{INFN, Sezione di Bari, I-70125, Bari, Italy}
\author{Václav Poto\v{c}ek}
\affiliation{Department of Physics, Faculty of Nuclear Sciences and Physical Engineering, Czech Technical University in Prague, B\v{r}ehov\'a 7, 115 19 Praha 1-Star\'e M\v{e}sto, Czech Republic}
\author{Martin \v{S}tefa\v{n}\'ak}
\affiliation{Department of Physics, Faculty of Nuclear Sciences and Physical Engineering, Czech Technical University in Prague, B\v{r}ehov\'a 7, 115 19 Praha 1-Star\'e M\v{e}sto, Czech Republic}
\author{Francesco V. Pepe}
\affiliation{Dipartimento Interuniversitario di Fisica, Universit\`a di Bari, I-70126 Bari, Italy}
\affiliation{INFN, Sezione di Bari, I-70125, Bari, Italy}
% \affil{$^*$corresponding author, e-mail: vaclav.potocek@fjfi.cvut.cz}

\date{today}

\begin{abstract}
Absorption of two-state coined quantum walks on a finite line  with two sinks located at $N$ and $-N$ is investigated. Elaborating on the results of Konno et al., J. Phys. A: Math. Gen. {\bf 36} 241 (2003), we derive closed formulas for the absorption probabilities at the boundaries in the limit of large system size $N$. Two limiting cases are considered, with the starting position $k$ being independent of $N$, or kept at a constant distance $\delta$ from one of absorbers. In the first scenario, the absorption probability is determined only by the coin parameter and polar angle of the initial coin state decomposed into the eigenbasis of the coin operator. In the second case, a correction depending exponentially on $\delta$ is introduced. Finally, we perform an extensive numerical investigation for small system size $N$, showing excellent agreement between numerical and analytical results. 
\end{abstract}

\maketitle

\section{Introduction}

Quantum walks \cite{Aharonov1993,Meyer1996,Farhi1998} have become a fundamental concept in the study of quantum dynamics and quantum information processing \cite{Ambainis2003}, see e.g. \cite{kadian2021quantum} for a recent overview. Indeed, they offer rich potential developing search algorithms \cite{Childs2004,Shenvi2003,Potocek2009,reitzner2009,Lovett2012,janmark2014,meyer2015,chakraborty2016,chakraborty2020,chakraborty_finding_2020,ambainis_quadratic_2020,apers_unified_2021,apers_quadratic_2022} and universal quantum computation \cite{childs2009universal,Lovett2010,Childs2013}. Moreover, they form a useful testbed for probing the interplay between coherence and measurement when studying phenomena like hitting times ~\cite{kempe:hitting:2005,krovi2006a,Krovi:2006,yin:hitting:223,wang:hitting:2024}, first passage \cite{friedman:passage:2017,thiel:arrival:2018,liu:transition:2020} or recurrence \cite{Grunbaum2013,gruenbaum2014,carvalho:recurrence:2017,grunbaum:2018,grunbaum:recurrence:2020,jacq:recurrence:2021,stefanak_rec_2023,stefanak:rec:2025}.

\newtext{Quantum walks have also been intensively studied as models of coherent transport on graphs and networks \cite{Ambainis2001,Mulken2011}. In this context, absorbing vertices or sinks provide a natural way to model irreversible detection, loss, or transfer to an external environment. Absorption problems are closely related to hitting times, first-passage processes, and recurrence, since they probe how repeated monitoring modifies the coherent spreading of the walker. For one-dimensional quantum walks, the dependence of the absorption probability on the initial condition for a finite line with two sinks was analyzed in \cite{Konno:absorp:2003}, while the time dependence of absorption probability for quantum walks with one and two sinks was studied in \cite{bach:absorp}, where an extension to $d$-dimensional walks with $(d-1)$-dimensional absorbing wall was also investigated. Absorbing times for quantum walks on the hypercube were studied in \cite{yamasaki:absorp}. Related variants include one-dimensional quantum walks with a moving boundary \cite{KwekSetiawan2011} and quantum walks with one variable absorbing boundary \cite{Wang2017}. More recently, absorption-based qubit estimation in discrete-time quantum walks was investigated in \cite{Amorim2026}.

More generally, transport in quantum walks can be strongly affected by localization phenomena. For example, in Grover-type walks, spectral degeneracies may lead to trapping, where part of the probability remains localized near the initial vertex \cite{Inui2005,falkner:limit:2014,stefanak2014,machida:limit:2015,stefanak_continuous_2012,kollar_complete_2020,stefanak_stability_2014,watabe_limit_2008,KonnoSegawaStefanak,Segawa_2023}. 
Here, and in the following, by trapping we mean linear spectral trapping associated with localized eigenstates or dark states of the unitary quantum-walk evolution. This should be distinguished from nonlinear self-trapping induced by Kerr-type intensity-dependent phase shifts, where localization originates from a nonlinear feedback mechanism rather than from point-spectrum degeneracy \cite{Buarque2020}, or from Anderson localization, which is a consequence of static spatial disorder in one-dimensional \cite{Joye2010,ahlbrecht_disordered_2011,Ahlbrecht2011,Ahlbrecht2012,Cedzich2013} and certain higher-dimensional models \cite{Joye2012,schaefer_dynamical_2025}.
Such effects are relevant for absorption problems because they can reduce the total probability reaching the sinks. Dynamical percolation provides one mechanism for suppressing trapping and restoring transport efficiency \cite{kollar_asymptotic_2012,kollar_discrete_2014,stefanak:2016,mares2019,mares2020,mares:2020b,mares:2022}, while related questions have also been studied in continuous-time quantum-walk models \cite{Mulken2007,Mulken2007b,Schijven_2012,darasz:fractals:2014}.
The investigation of different aspects of quantum-walk dynamics with absorption is also motivated by their experimental relevance, supported by implementations in trapped atoms or ions \cite{karski2009quantum,zahringer2010realization} and photonic quantum systems \cite{kitagawa2012observation,crespi2013anderson,peruzzo2010quantum,sansoni2012twoparticle,benedetti2021quantum,yang2024programmable,sansoni2025noisy,stefanutti2026implementation,schreiber2010photons,Schreiber2011,Schreiber2012,Nitsche2018,Chen2024}, where controlled losses or out-coupling can emulate absorbing processes.}

In the present paper we focus on a two-state discrete-time quantum walk with a one-parameter coin on a finite line with absorbing sinks on positions $\pm N$. This model was investigated in \cite{Konno:absorp:2003} where the authors utilized combinatorial approach (path counting and generating function methods) to derive absorption probabilities on the left and right end of the line in dependence of the coin and the initial state. However, the formulas in \cite{Konno:absorp:2003} are implicit - they involve integrals of functions which have to be determined from recurrence relations for a particular case given by the length of the line and the position of the initial vertex. Elaborating on the results of \cite{Konno:absorp:2003}, we show that the integrands can be cast in a form suitable for asymptotic analysis in the limit of large $N$. We focus on two limits, where either the initial position $k$ is fixed as $N\to\infty$, or $k$ is close to one of the absorbing boundary (i.e. $N-k = \delta$ or $N+k = \delta$ is constant as $N\to\infty$). Utilizing the two-scale convergence analysis, we derive closed formulas for the absorption probabilities. For further simplification we decompose the initial state into the coin eigenstates, a useful trick which can highlight otherwise hidden features of quantum walks \cite{Stefanak2011,stefanak2014,bezdekova:2015,stefanak:2016b}. We show that for a fixed starting position $k$ the absorption probability is determined only by the coin parameter and the polar angle of the initial coin state. Considering finite distance $\delta$ from one of the absorbers yields a correction which exponentially decreases with $\delta$. Finally, inspired by the results obtained in the asymptotic limit we propose an approximation of the absorption probability for finite $N$ and $k$.

The rest of the paper is organized as follows: In Section~\ref{sec:overview} we introduce the model and review the previously derived results from the literature. In Section~\ref{sec:parity} we discuss the parity symmetry which allows to restrict the analysis to starting positions $0\leq k\leq N-1$.  We evaluate the absorption probabilities in the asymptotic limit of large system size $N\to\infty$ in Section \ref{sec:analytical}. In Section~\ref{sec:numerical} we \newtext{propose an approximation of absorption probability for finite $N$ and perform an extensive numerical investigation of small size systems, showing exponentially fast convergence to the analytical results.} We conclude and present an outlook in Section~\ref{sec:conclusions}. More technical details \newtext{of the derivations} are left for Appendices \ref{sec:integrands} and \ref{app:b}. \newtext{Finally, in Appendix \ref{app:c} we show that the results derived in the main text are easily modified for lines of even lengths, i.e. with absorbers at $-N$ and $N+1$.}

\section{Notation and overview of the existing results}
\label{sec:overview}

\newtext{In this section we first introduce the finite-line quantum-walk model studied in this work and then recall the absorption-probability formulas from Ref.~\cite{Konno:absorp:2003}, which provide the starting point for the asymptotic analysis developed in the following sections.

\subsection{Finite-line quantum-walk model}}
We consider the propagation of a quantum walker on a finite discrete line with absorbing barriers (sinks) located at both ends. The vertices of this line are labeled from $-N$ to $N$ ($N\geq 2$), with absorbing sinks placed at vertices $-N$ and $N$.
The Hilbert space for the quantum walk is defined as a tensor product:
   \begin{equation}
\mathcal{H} = \mathcal{H}_P \otimes \mathcal{H}_C,
\end{equation}
where $\mathcal{H}_P$ is the position Hilbert space spanned by basis vectors $|m\rangle$, $m = -N, \dots, N$, representing the position of the quantum walker. The coin space $\mathcal{H}_C$ describes the internal degree of freedom (coin state) of the walker, which in our case is a two-dimensional space spanned by vectors $|L\rangle$ and $|R\rangle$ corresponding to left and right directions, respectively.

Initially, the walker is placed at position $k$ ranging from $-N+1$ to $N-1$, with an initial coin state $|\psi_c\rangle \in \mathcal{H}_C$ given by
\begin{equation}
\label{init:coin:stb}
    \ket{\psi_c} = a\ket{L} + b \ket{R}, \quad |a|^2 + |b|^2 = 1. 
\end{equation}
Each discrete-time step of the quantum walk consists of two operations: a coin flip followed by a shift. The evolution operator per step is given by:
\begin{equation}
\hat{U} = \hat{S}(\hat{I}_P \otimes \hat{C}),
\end{equation}
where the shift operator $\hat{S}$ conditionally propagates the walker left or right depending on the coin state
\begin{equation}
\hat{S} = \sum_{x=-N}^{N} \big(|x-1\rangle\langle x| \otimes |L\rangle\langle L| + |x+1\rangle\langle x| \otimes |R\rangle\langle R|\big),
\end{equation} 
where we consider periodic boundary condition, i.e. $\ket{N+1}\equiv\ket{-N}$ and $\ket{-N-1}\equiv\ket{N}$. However, due to the absorption at vertices $N$ and $-N$ described later, the walker never remains in states $\ket{N}$ or $\ket{-N}$ before $\hat S$ is applied.
The coin operator $\hat{C}$ acts only on the coin space $\mathcal{H}_C$.
In our study, we only consider a one-parameter family of coin matrices \begin{equation}
\label{coin}
\hat{C}(\theta) =
\begin{pmatrix}
\cos \theta & \sin \theta \\
\sin \theta & -\cos \theta
\end{pmatrix}, \quad \theta \in (0,\frac{\pi}{2}).
\end{equation}
In principle, the coin could be an arbitrary $\rm U(2)$ operator. However, the global phase has no physical significance and the relative phases between coin matrix elements were shown to be largely irrelevant for the two-state quantum walk models \cite{Tregenna2003,goyal_2015_unitary_equiv}, as they can be compensated by the relative phase between the amplitudes of the initial coin state (\ref{init:coin:stb}).
Namely, any $\rm U(2)$ operator can be written in the form
\begin{equation}
\label{coin-abc}
\hat C(\theta, \alpha, \beta, \gamma) = \mathop{\mathrm{diag}}(e^{i \alpha}, e^{i \beta})
\hat C(\theta) \exp\left(i \gamma \sigma_z \right)
\end{equation}
due to a slight modification of Pauli decomposition, but it is easily shown that a quantum walk with coin $\hat C(\theta, \alpha, \beta, \gamma)$ and initial coin state $\ket{\psi_c}$ leads to the same probability distribution in position in every step number as one with coin $\hat C(\theta)$ and initial coin state $\exp(-i \gamma \sigma_z) \ket{\psi_c}$. If $\ket{\psi_c}$ is allowed to be arbitrary in our model, then the restricted coin \eqref{coin} is without loss of generality.

In \eqref{coin}, the angle $\theta$ is a tunable parameter governing the effective spreading speed of the quantum walk \cite{miyazaki:2007,kempf:2009,bezdekova:2015}.
The special case $\theta = \pi/4$ corresponds to the well-known Hadamard coin, which has been extensively studied in the literature. We omit the trivial boundary cases $\theta=0,\, \frac{\pi}{2}$, which can be treated in a straightforward way. Indeed, for $\theta = 0$ the coin (\ref{coin}) is diagonal and the quantum walker propagates deterministically according to the amplitudes of the initial state (\ref{init:coin:stb}). Absorption probability on the left and right is given by $|a|^2$ and $|b|^2$, respectively. On the other hand, for $\theta=\frac{\pi}{2}$ the coin (\ref{coin}) becomes a Pauli $\sigma_x$ matrix and the quantum walker is bound to positions $k-1,k,k+1$. Unless the starting point is next to the absorbing barrier (i.e. $k=N-1$ or $k=-N+1$), the walker will not be absorbed.

We may describe the absorption using partial measurement of the walker's position after each step, continuing only if neither $N$ or $-N$ was measured, leading to a non-unitary evolution. Define projection operators
\begin{equation}
    \hat\Pi_L = |{-N}\rangle\langle -N| \otimes \hat{I}_C,
    \quad
    \hat\Pi_R = |N\rangle\langle N| \otimes \hat{I}_C
\end{equation}
and
\begin{equation}
    \hat{\Pi} = (\hat{I}_P - |N\rangle\langle N| - |{-N}\rangle\langle -N|) \otimes \hat{I}_C = I - \Pi_L - \Pi_R.
\end{equation}
For an initial state $\ket{\psi(0)} = \ket{k} \otimes \ket{\psi_c}$, the probability that the walker is detected in the left (right) absorbing boundary in step $t$, $t \ge 1$, is given by
\begin{equation}
    P_L^{(\psi_c)}(k,N,t) = \| \hat \Pi_L \hat U (\hat \Pi \hat U)^{t-1} \ket{\psi(0)} \|^2, \quad
    P_R^{(\psi_c)}(k,N,t) = \| \hat \Pi_R \hat U (\hat \Pi \hat U)^{t-1} \ket{\psi(0)} \|^2,
    \label{eq:pl,pr}
\end{equation}
so the effective (non-unitary) evolution operator is $\hat \Pi \hat U$. \newtext{Note that due to the bipartite nature of the dynamics of the discrete-time quantum walk considered here, either of the detection probabilities is trivially zero at times $t$ for which $t + k + N$ is an odd number, in addition to the restriction to the condition of finite spreading speed. Nevertheless, the asymptotic cumulative probabilities that we study in the following are not directly affected by either of these finite-time effects.}

Since for two-state quantum walks there are no dark states, i.e. localized eigenstates of the unperturbed evolution operator with no support on the sink vertices~\cite{KonnoSegawaStefanak}, the survival probability
\begin{eqnarray}
\label{eq:proj-dynamics}
    \|\psi(t)\|^2 = \|(\hat \Pi \hat U)^t \psi(0)\|^2 = 1 - \sum_{t'=1}^t \left( P_L^{(\psi_c)}(k,N,t') + P_R^{(\psi_c)}(k,N,t') \right)
\end{eqnarray}
tends to zero in the limit of large number of steps $t$. Hence, left and right absorption probabilities add up to 1 in the asymptotic limit
\begin{equation}
\label{pl:pr}
    \sum_{t=1}^\infty P_L^{(\psi_c)}(k,N,t) + \sum_{t=1}^\infty P_R^{(\psi_c)}(k,N,t) = P_L^{(\psi_c)}(k,N) + P_R^{(\psi_c)}(k,N) = 1.
\end{equation}
\newtext{\subsection{Known integral representation of the absorption probability}}
The asymptotic value of the left absorption probability $P_L^{(\psi_c)}(k,N)$ was studied in \cite{Konno:absorp:2003}, where it was shown that it can be expressed in the form
\begin{equation}
\label{abs:left}    P_L^{(\psi_c)}(k,N) = C_1(k,N) |a|^2+ C_2(k,N) |b|^2 + 2 \mathrm{Re}(C_3(k,N) \overline{a} b). 
\end{equation}
Here $a$ and $b$ are amplitudes of the initial coin state (\ref{init:coin:stb}) and the coefficients $C_j(k)$ are given by the integrals (see Theorem 2 from \cite{Konno:absorp:2003})
\begin{align}
\label{eq:cj:coef} C_1(k,N) &= \frac{1}{2\pi} \int_{0}^{2\pi} \left| \cos\theta\, p^{(2N)}_{N+k}(e^{i\phi}) + \sin\theta\, r^{(2N)}_{N+k}(e^{i\phi}) \right|^2\, d\phi, \\
\nonumber C_2(k,N) &= \frac{1}{2\pi} \int_{0}^{2\pi} \left| \sin\theta\, p^{(2N)}_{N+k}(e^{i\phi}) - \cos\theta\, r^{(2N)}_{N+k}(e^{i\phi}) \right|^2\, d\phi, \\
\nonumber C_3(k,N) &= \frac{1}{2\pi} \int_{0}^{2\pi}  \overline{\left( \cos\theta\, p^{(2N)}_{N+k}(e^{i\phi}) + \sin\theta\, r^{(2N)}_{N+k}(e^{i\phi}) \right)} \left(\sin\theta\, p^{(2N)}_{N+k}(e^{i\phi}) - \cos\theta\, r^{(2N)}_{N+k}(e^{i\phi}) \right) d\phi.
\end{align}
We kept the notation $p_{i}^{(j)}$, $r_{i}^{(j)}$ used in \cite{Konno:absorp:2003}, where the lower index $i$ indicates the starting point of the walk, while the upper index $j$ denotes the position of the right absorbing barrier. Note that \cite{Konno:absorp:2003} considers the left sink to be located at 0. Our configuration with sinks at $\pm N$ and initial vertex $k$ is equivalent to $i=N+k$ and $j=2N$. The explicit form of the functions $p^{(2N)}_{N+k}(z)$ and $r^{(2N)}_{N+k}(z)$ read
\begin{align}
    \nonumber p^{(2N)}_{N+k}(z) & = \left(\frac{z}{2} + E_z\right) \lambda_+^{N+k-1} + \left(\frac{z}{2} - E_z\right) \lambda_-^{N+k-1} = \frac{z}{2} x_{N+k-1} + E_z y_{N+k-1}, \\
\label{eq:pn:rn}    r^{(2N)}_{N+k}(z) & = (-1)^{N-k} C_z \left(\lambda_+^{N-k-1} - \lambda_-^{N-k-1}\right) = (-1)^{N-k} C_z y_{N-k-1},
\end{align}
where $\lambda_\pm$ are given by
\begin{equation}
\label{eq:lambda}
\lambda_{\pm} = \frac{ z^2 - 1 \pm \sqrt{1 + z^4 + 2 \cos(2\theta)z^2}}{2\cos\theta z},
\end{equation}
and $C_z$ and $E_z$ are determined from the equations
\begin{align}
    \nonumber C_z\, y_{2N-2} & = z \cos\theta\, C_z\, y_{2N-3} - z \sin\theta\, \left(\frac{z}{2}\, x_1 + E_z\, y_1\right) , \\
\label{eq:cz:ez}    C_z\, y_1 & = z \sin\theta\, \left(\frac{z}{2}\, x_{2N-2} + E_z\, y_{2N-2}\right).
\end{align}
Here we have introduced
\begin{align}
 x_n &= \lambda_{+}^n + \lambda_{-}^n, \quad y_n = \lambda_{+}^n - \lambda_{-}^n,
 \label{eq:x,y}
\end{align}
to simplify the notation of \cite{Konno:absorp:2003}.

In Appendix~\ref{sec:integrands} we show that the integrands of coefficients $C_j(k,N)$ can be reduced to the following form
\begin{equation}
\begin{aligned}
    I_1(\phi;k,N) &=  \frac{1 - \cos^2\theta \cos(2\alpha) - \sin^2\theta \cos((2N-2k-2)\alpha)}{1 - \cos^2\theta \cos(2\alpha) - \sin^2\theta \cos((4N-2)\alpha)}, \\
     I_2(\phi;k,N) &= \frac{\sin^2\theta (1 - \cos((2N-2k)\alpha))}{1 - \cos^2\theta \cos(2\alpha) - \sin^2\theta \cos((4N-2)\alpha)}, \\
    \mathrm{Re}\ I_3(\phi;k,N) &= \frac{\frac12\sin\theta\cos\theta (1 - \cos(2\alpha) - \cos((2N-2k)\alpha) + \cos((2N-2k-2)\alpha))}{1 - \cos^2\theta \cos(2\alpha) - \sin^2\theta \cos((4N-2)\alpha)} .
    \label{eq:i1}
\end{aligned}
\end{equation}
where
\begin{equation}
    \alpha = \arccos \frac{\sin\phi}{\cos\theta}.
    \label{eq:alpha0}
\end{equation}
Note that the imaginary part of $I_3$ is an odd function of $\phi$, so its integral over $(0,2\pi)$ vanishes. Hence, $C_3(k,N)$ is real-valued.

\section{Parity symmetry}
\label{sec:parity}

In the following, we will actively use the parity symmetry of the system, \newtext{represented by the operator}
\begin{equation}
    \hat P: \ket{x} \otimes \ket{c} \mapsto \ket{{-x}} \otimes (\sigma_y \ket{c}).
\label{eq:parity-op}
\end{equation}
\newtext{We emphasize that this is a symmetry of the dynamics, in the sense that it maps an evolution of the system with arbitrary initial state to another valid instance (up to a global phase in each step), as shown below. Individual initial states may or may not be invariant under $\hat P$ -- a state which is, remains an eigenstate of $\hat P$ during its evolution, leading to a reflection symmetry of its probability distribution of position, and equal absorption probabilities in the left and right boundaries.}

It is easy to show that $\hat P$ commutes with the shift operator $\hat S$. For the coin,
\begin{equation}
    \hat P \hat C(\theta) \hat P = -\hat C(\theta),
\end{equation}
but this only induces a global phase.
%, the probability distribution of positions is not affected.
Finally,
\begin{equation}
    \hat P \hat \Pi_L \hat P = \Pi_R, \quad
    \hat P \hat \Pi_R \hat P = \Pi_L, \quad
    \hat P \hat \Pi \hat P = \Pi.
\end{equation}
These properties directly imply, by the definitions \eqref{eq:pl,pr},
\begin{equation}
    P_R^{(\tilde{\psi}_c)}(-k,N)
    =
    P_L^{(\psi_c)}(k,N)
    \label{eq:pr-pl-symm}
\end{equation}
for
\begin{equation}
    \ket{\smash{\tilde{\psi}_c}}
    =
    \exp\left(-i\frac{\pi}{2}\sigma_z\right)
    \sigma_x\ket{\psi_c}
    =
    \sigma_y\ket{\psi_c}
    =
    \corr{-ib\ket{L}+ia\ket{R}}.
\end{equation}
Thus,
\begin{equation}
    P_R^{(\tilde{\psi}_c)}(-k,N)
    =
    1-P_L^{(\tilde{\psi}_c)}(-k,N)
    =
    1-C_1(-k,N)|b|^2
    -C_2(-k,N)|a|^2
    +2\mathrm{Re}
    \left(
        C_3(-k,N)\overline{b}a
    \right).
\end{equation}
Comparing the linearly independent terms, we obtain the identities
\begin{equation}
    C_{1,2}(-k,N)
    =
    1-C_{2,1}(k,N),
    \qquad
    C_3(-k,N)
    =
    C_3(k,N).
    \label{eq:parity}
\end{equation}

\section{Analytical derivation for large size $N$}
\label{sec:analytical}

In the following we will evaluate the integrals $C_j(k,N)$ (\ref{eq:cj:coef}) in the limit $N \to \infty$. Note that by \eqref{eq:i1}, the integrands all have the form of a linear combination of terms $\cos(2m\alpha), m \in \mathbb{N}_0$, divided by a common denominator.

We now restrict $k$ to $0 \le k < N$, as the other half of initial positions follows from \eqref{eq:parity}. In this case, all the multiplicators $m$ satisfy $m < N$. On the other hand, the denominator has a term $\cos((4N-2)\alpha)$. This observation will come useful in two respects.

Consider separately the cases where $\alpha$ is real and where it is imaginary, i.e., where $|\sin\phi| \le \cos\theta$ or $|\sin\phi| > \cos\theta$, respectively. In the latter case, we equivalently replace
\begin{equation}
    \frac{\cos(2m\alpha)}{1 - \cos^2\theta \cos(2\alpha) - \sin^2\theta \cos((4N-2)\alpha)}
\end{equation}
by
\begin{equation}
    \frac{\cosh(2m\mu)}{1 - \cos^2\theta \cosh(2\mu) - \sin^2\theta \cosh((4N-2)\mu)},
\end{equation}
where $\mu = \Im \alpha$ (note that for a particular choice of branches of $\arccos$, $\alpha$ can be written as $i \mu$ or $\pi + i \mu$).
However, now we can see that the denominator is $\Theta(e^{4N\mu})$ while the numerator is $\Theta(e^{2m\mu})$ with $m < N$. Thus the pointwise limit as $N \to \infty$ is zero over the entire region of $\phi$ where $|\sin\phi| > \cos\theta$, and it's easy to show by argument of dominated convergence that in turn the corresponding part of each of the three integrals vanishes.

For $|\sin\phi| < \cos\theta$, the function has an oscillating character in the auxiliary variable $\alpha$. We start by substituting for $\alpha$ as the integration variable. Over the three subintervals where $|\sin\phi| < \cos\theta$, i.e.,
\begin{equation}
    \{\phi \in (0,2\pi) \mid |\sin\phi| < \cos\theta\}
    = \left(0,\frac\pi2 - \theta\right) \cup \left(\frac\pi2 + \theta, \frac{3\pi}{2} - \theta\right) \cup \left(\frac{3\pi}{2} + \theta, 2\pi\right),
\end{equation}
$\cos\alpha$ goes from $0$ to $1$, then from $1$ to $-1$, and finally from $-1$ to $0$, and the sign of $\sin\alpha$ does not enter the integrands, so $\alpha$ can be taken to cover a single interval of length $2\pi$. The Jacobian of the substitution is
\begin{equation}
    \left|\frac{d\phi}{d\alpha}\right| = \frac{\cos\theta|\sin\alpha|}{\cos\phi} = \frac{\cos\theta|\sin\alpha|}{\sqrt{1-\cos^2\theta\cos^2\alpha}}.
\end{equation}
With this change the integrands $I_j(\phi;k,N)$ become
\begin{equation}
\begin{aligned}
    \tilde I_1(\alpha;k,N) &= \frac{1 - \cos^2\theta \cos(2\alpha) - \sin^2\theta \cos((2N-2k-2)\alpha)}{1 - \cos^2\theta \cos(2\alpha) - \sin^2\theta \cos((4N-2)\alpha)} \frac{\cos\theta|\sin\alpha|}{\sqrt{1-\cos^2\theta\cos^2\alpha}}, \\
    \tilde I_2(\alpha;k,N) &= \frac{\sin^2\theta (1 - \cos((2N-2k)\alpha))}{1 - \cos^2\theta \cos(2\alpha) - \sin^2\theta \cos((4N-2)\alpha)} \frac{\cos\theta|\sin\alpha|}{\sqrt{1-\cos^2\theta\cos^2\alpha}}, \\
    \Re \tilde I_3(\alpha;k,N) &= \frac{\frac12\sin\theta\cos\theta (1 - \cos(2\alpha) - \cos((2N-2k)\alpha) + \cos((2N-2k-2)\alpha))}{1 - \cos^2\theta \cos(2\alpha) - \sin^2\theta \cos((4N-2)\alpha)} \frac{\cos\theta|\sin\alpha|}{\sqrt{1-\cos^2\theta\cos^2\alpha}}\\
    \label{eq:i1,2,3}
\end{aligned}
\end{equation}
and are in ideal shape for two-scale convergence analysis. For $\theta \in (0, \pi/2)$, the functions are $L^\infty$, satisfy all assumptions of the Nguetseng-Allaire theory \cite{Nguetseng:2scaleconv:1989,Allaire:2scaleconv:1992} and two-scale converge to
\begin{equation}
\begin{aligned}
    \tilde I_1(\alpha,\beta;k) &= \frac{1 - \cos^2\theta \cos(2\alpha) - \sin^2\theta \cos(\beta - (2k+1)\alpha)}{1 - \cos^2\theta \cos(2\alpha) - \sin^2\theta \cos(2\beta)} \frac{\cos\theta|\sin\alpha|}{\sqrt{1-\cos^2\theta\cos^2\alpha}}, \\
    \tilde I_2(\alpha,\beta;k) &= \frac{\sin^2\theta (1 - \cos(\beta-(2k-1)\alpha))}{1 - \cos^2\theta \cos(2\alpha) - \sin^2\theta \cos(2\beta)} \frac{\cos\theta|\sin\alpha|}{\sqrt{1-\cos^2\theta\cos^2\alpha}}, \\
    \Re \tilde I_3(\alpha,\beta;k) &= \frac{\frac12\sin\theta\cos\theta (1 - \cos(2\alpha) - \cos(\beta-(2k-1)\alpha) + \cos(\beta-(2k+1)\alpha))}{1 - \cos^2\theta \cos(2\alpha) - \sin^2\theta \cos(2\beta)} \frac{\cos\theta|\sin\alpha|}{\sqrt{1-\cos^2\theta\cos^2\alpha}}\\
    \label{eq:i1,2,3-beta}
\end{aligned}
\end{equation}
as $N \to \infty$ (for test functions in $(\alpha, (2N-1)\alpha)$). The two-scale convergence then gives
\begin{eqnarray}
\nonumber   \lim_{N \to \infty} \frac{1}{2\pi} \int_0^{2\pi} \tilde I(\alpha; k,N) d\alpha
    & = & \lim_{N \to \infty} \frac{1}{2\pi} \int_0^{2\pi} \tilde I(\alpha, (2N-1)\alpha; k) d\alpha \\
    & = & \frac{1}{(2\pi)^2} \int_0^{2\pi} \int_0^{2\pi} \tilde I(\alpha, \beta; k) d\alpha d\beta .
\end{eqnarray}
For $k = \mathrm{const}.$, this leads to the integrals
\begin{equation}
\begin{aligned}
    \frac{1}{2\pi} \int_0^{2\pi} \frac{1}{a - \sin^2\theta \cos (2\beta)} d\beta
    &= \frac{1}{\sqrt{(a-\sin^2\theta)(a+\sin^2\theta)}}
    = \frac{1}{2\cos\theta|\sin\alpha| \sqrt{1-\cos^2\theta\cos^2\alpha}}, \\
    \frac{1}{2\pi} \int_0^{2\pi} \frac{\cos(\beta + x)}{a - \sin^2\theta \cos (2\beta)} d\beta
    &= 0, \quad x = \mathrm{const.} \in \mathbb{R}
\end{aligned}
\end{equation}
where
\begin{equation}
    a = 1 - \cos^2\theta \cos(2\alpha),
\end{equation}
thus
\begin{equation}
\begin{aligned}
    C_1 \coloneq \lim_{N \to \infty} C_1(k,N) &= \frac{1}{2\pi} \int_0^{2\pi}
    \frac{1 - \cos^2\theta \cos(2\alpha)}{2 (1-\cos^2\theta\cos^2\alpha)} d\alpha, \\
    C_2 \coloneq \lim_{N \to \infty} C_2(k,N) &= \frac{1}{2\pi} \sin^2\theta \int_0^{2\pi}
    \frac{1}{2(1-\cos^2\theta\cos^2\alpha)} d\alpha, \\
    C_3 \coloneq \lim_{N \to \infty} C_3(k,N) &= \frac{1}{2\pi} \frac{\sin\theta\cos\theta}{2} \int_0^{2\pi}
    \frac{1 - \cos(2\alpha)}{2(1-\cos^2\theta\cos^2\alpha)} d\alpha, 
    \label{eq:c123-avg}
\end{aligned}
\end{equation}
are independent of $k$. The only case in which the $k$-dependent terms can contribute in the limit $N \to \infty$ is when we consider the initial position $k$ dependent on $N$ and fix $N-k = \delta = \mathrm{const.} > 0$ within the limit. In this setting,
\begin{equation}
\begin{aligned}
    C_1(\delta) \coloneq \lim_{N \to \infty} C_1(N-\delta,N) &= \frac{1}{2\pi} \int_0^{2\pi}
    \frac{1 - \cos^2\theta \cos(2\alpha) - \sin^2\theta \cos((2\delta-2)\alpha)}{2(1-\cos^2\theta\cos^2\alpha)} d\alpha, \\
   C_2(\delta) \coloneq \lim_{N \to \infty} C_2(N-\delta,N) &= \frac{1}{2\pi} \sin^2\theta \int_0^{2\pi}
    \frac{1 - \cos(2\delta\alpha)}{2(1-\cos^2\theta\cos^2\alpha)} d\alpha, \\
    C_3(\delta) \coloneq \lim_{N \to \infty} C_3(N-\delta,N) &= \frac{1}{2\pi}\frac{\sin\theta\cos\theta}{2} \int_0^{2\pi}
    \frac{1 - \cos(2\alpha) - \cos(2\delta\alpha) + \cos((2\delta-2)\alpha)}{2(1-\cos^2\theta\cos^2\alpha)} d\alpha. \\
    \label{eq:c123-avg-delta}
\end{aligned}
\end{equation}

We see that (\ref{eq:c123-avg}) and (\ref{eq:c123-avg-delta}) can be expressed as linear combinations of integrals of the form
\begin{equation}
\label{eq:Am}
    A_m = \frac{1}{2\pi} \int_0^{2\pi} \frac{\cos(2m\alpha)}{2 (1-\cos^2\theta\cos^2\alpha)} d\alpha.
\end{equation}
We study these in Appendix \ref{app:b} where we show that
\begin{eqnarray}
  A_m = \frac{1}{2\sin\theta} q^m, \quad q = \frac{1-\sin\theta}{1+\sin\theta}.
\end{eqnarray}
With this, evaluating the forms (\ref{eq:c123-avg}) and (\ref{eq:c123-avg-delta}) becomes a matter of plugging in:
\begin{equation}
\begin{aligned}
    C_1 &= A_0 - \cos^2\theta A_1 = 1 - \frac{\sin\theta}{2}, \\
    C_2 &= \sin^2\theta A_0 = \frac{\sin\theta}{2}, \\
    C_3 &= \frac{\sin\theta\cos\theta}{2} (A_0 - A_1) = \frac{\sin\theta(1-\sin\theta)}{2\cos\theta}, \\
    \label{eq:result-const}
\end{aligned}
\end{equation}
and similarly
\begin{equation}
\begin{aligned}
    C_1(\delta) &= 1 - \frac{\sin\theta}{2}\left(1 + q^{\delta - 1}\right), \\
    C_2(\delta) &= \frac{\sin\theta}{2}\left(1 - q^{\delta-1}\right), \\
    C_3(\delta) &= \frac{\sin\theta(1-\sin\theta)}{2\cos\theta} \left(1 +  q^{\delta - 1}\right).
    \label{eq:result-n-delta}
\end{aligned}
\end{equation}

Through the symmetry \eqref{eq:parity}, the validity of \eqref{eq:result-const} extends to negative $k$ values. The last case of practical interest is the behaviour when the initial position is fixed near the left absorbing barrier rather than the right. Here, \eqref{eq:parity} gives
\begin{equation}
\begin{aligned}
    C_1(-\delta) \coloneq \lim_{N \to \infty} C_1(-N+\delta,N) &= 1 - \frac{\sin\theta}{2}\left(1 - q^{\delta - 1}\right), \\
    C_2(-\delta) \coloneq \lim_{N \to \infty} C_2(-N+\delta,N) &= \frac{\sin\theta}{2} \left(1 + q^{\delta - 1}\right), \\
    C_3(-\delta) \coloneq \lim_{N \to \infty} C_3(-N+\delta,N) &= \frac{\sin\theta(1-\sin\theta)}{2\cos\theta}\left( 1 + q^{\delta - 1}\right).
    \label{eq:result--n+delta}
\end{aligned}
\end{equation}

To further simplify the absorption probability (\ref{abs:left}) we express it in a matrix form
\begin{equation}
    P_L^{(\psi_c)}(k,N) = \begin{pmatrix}
        \bar a & \bar b
    \end{pmatrix} Q_L(k,N) \begin{pmatrix}
        a \\ b
    \end{pmatrix}, \quad
    Q_L(k,N) = \begin{pmatrix}
        C_1(k,N) & C_3(k,N) \\ C_3(k,N) & C_2(k,N)
    \end{pmatrix}
\end{equation}
In the limit considered here,
\begin{equation}
    \lim_{N \to \infty} Q_L(k,N) = \begin{pmatrix}
        \frac12 & 0 \\ 0 & \frac12
    \end{pmatrix} + \frac{1-\sin\theta}{2\cos\theta} \begin{pmatrix}
        \cos\theta & \sin\theta \\ \sin\theta & -\cos\theta
    \end{pmatrix}
    = \frac12 I + \frac{1-\sin\theta}{2\cos\theta} \hat C(\theta)
\end{equation}
\newtext{and thus the limit absorption probability at the left end of the line reads
\begin{equation}
    P_L^{(\psi_c)} \coloneq \frac12+\frac{1-\sin\theta}{2\cos\theta} \langle \psi_c|\hat C(\theta)|\psi_c\rangle .
    \label{eq:pl:matrix}
\end{equation}
This admits further simplification in the eigenbasis of the matrix $\hat C(\theta)$,}
\begin{align}
\ket{\theta^+}  = \cos\frac{\theta}{2} \ket{L} + \sin\frac{\theta}{2} \ket{R}, \quad \ket{\theta^-} & = -\sin\frac{\theta}{2} \ket{L} + \cos\frac{\theta}{2} \ket{R} 
\end{align}
such that $\hat C(\theta) \ket{\theta^\pm} = \pm \ket{\theta^\pm}$. Writing the initial coin state as
\begin{equation}
\label{init:eigenbasis}
    \ket{\psi_c} = u \ket{\theta^+} + v \ket{\theta^-} = \cos\frac{\rho}{2} \ket{\theta^+} +\sin\frac{\rho}{2} e^{i\varphi} \ket{\theta^-}, 
\end{equation}
where $\rho\in\langle0,\pi\rangle$ and $\varphi\in\langle 0,2\pi)$ are the polar and the azimuthal angles in the Bloch sphere representation, \newtext{\eqref{eq:pl:matrix} gives}
\begin{equation}
 \label{eq:pl}
 P_L^{(\psi_c)} = \frac12 + \frac{1-\sin\theta}{2\cos\theta} (|u|^2 - |v|^2) = \frac12 + \frac{1-\sin\theta}{2\cos\theta} \cos\rho ,
\end{equation}
notably insensitive of the relative phase $\varphi$. The absorption probability is thus determined by the coin parameter $\theta$ and the polar angle $\rho$ of the initial coin state $\ket{\psi_c}$. The absorption probability at the right end of the line is the complement to unity,
\begin{equation}
\label{eq:pr}
P_R^{(\psi_c)} \coloneq \lim_{N\to\infty} P_R^{(\psi_c)}(k,N) = \frac12 - \frac{1-\sin\theta}{2\cos\theta}\cos\rho.
\end{equation}
The sensitivity of $P_L^{(\psi_c)}$, $P_R^{(\psi_c)}$ on the initial state is determined by the coin parameter. For the Hadamard walk ($\theta = \pi/4$), the minimum and maximum probabilities, achieved by the two coin eigenstates, are
\begin{equation}
    P_L^{(\theta^\pm)} = \ P_R^{(\theta^\mp)} = \frac12 \pm \frac{\sqrt2 - 1}{2},
\end{equation}
or $1/\sqrt{2}$ and $1-1/\sqrt{2}$. In the case $\theta = 0$, the coin eigenbasis is the standard basis so the difference of the two expressions vanishes. We obtain
\begin{equation}
    \lim_{N \to \infty} Q_L(k,N) = \begin{pmatrix}
        1 & 0 \\ 0 & 0
    \end{pmatrix},
\end{equation}
leading to $P_L^{(\psi_c)} = |a|^2$, consistent with the earlier discussion of this special case. For the other exclusion $\theta = \frac{\pi}{2}$ the expressions (\ref{eq:pl}) and (\ref{eq:pr}) obtained in this section are not valid.

\begin{figure}
    \centering
    \includegraphics[width=0.6\linewidth]{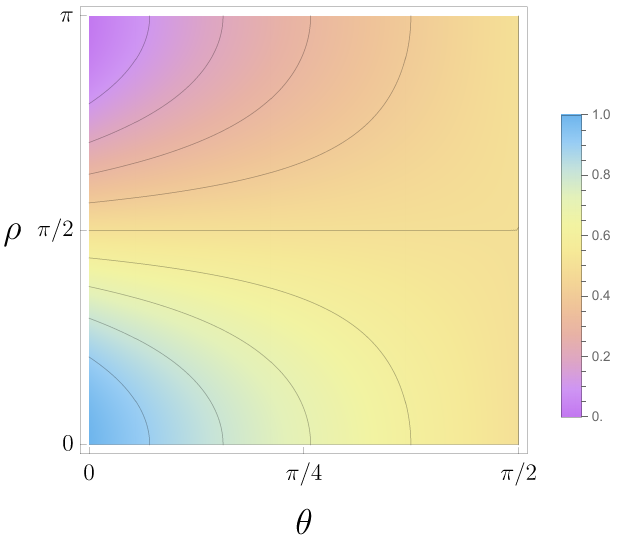}
    \caption{Absorption probability at the left end of the line (\ref{eq:pl}) as a function of the coin angle $\theta$ and the polar angle $\rho$ of the initial coin state. The lines show the contours $P_L = j/10$ for $j=1,\ldots 9$.
    }
    \label{fig:abs:left:dens}
\end{figure}

We illustrate these results in Figure~\ref{fig:abs:left:dens} where we show the absorption probability at the left end (\ref{eq:pl}) as a function of the coin angle $\theta$ and the polar angle $\rho$.

The distribution of the absorption probabilities does not vary in $k$ (in the limit $N \to \infty$) as long as $k$ does not scale with $N$. As obtained above, their values will start showing position dependence if $k = N - \mathrm{const.}$ or $k = -N + \mathrm{const}$. In the matrix form as above,
\begin{equation}
\begin{aligned}
    \lim_{N \to \infty} Q_L(N - \delta,N) &= \frac12 I + \frac{1-\sin\theta}{2\cos\theta} \hat C(\theta) + q^{\delta-1} \frac{\sin\theta}{2(1+\sin\theta)} \begin{pmatrix}
        -1-\sin\theta & \cos\theta \\ \cos\theta & -1+\sin\theta
    \end{pmatrix}, \\
    \lim_{N \to \infty} Q_L(-N + \delta,N) &= \frac12 I + \frac{1-\sin\theta}{2\cos\theta} \hat C(\theta) + q^{\delta-1} \frac{\sin\theta}{2(1+\sin\theta)} \begin{pmatrix}
        1-\sin\theta & \cos\theta \\ \cos\theta & 1+\sin\theta
    \end{pmatrix}. \\
    \label{eq:ql-n-delta}
\end{aligned}\end{equation}
These matrices also simplify if cast in the coin matrix eigenbasis $(\ket{\theta^+}, \ket{\theta^-})$. In this basis, they become
\begin{equation}
\begin{aligned}
    \lim_{N \to \infty} \tilde Q_L(N - \delta,N) &= 
    \frac12 \begin{pmatrix}
        1 & 0 \\ 0 & 1
    \end{pmatrix} + \frac{1-\sin\theta}{2\cos\theta} \begin{pmatrix}
        1 & 0 \\ 0 & -1
    \end{pmatrix} + q^{\delta-1} \frac{\sin\theta}{2(1+\sin\theta)} \begin{pmatrix}
        -1 & 1 \\ 1 & -1
    \end{pmatrix}, \\
    \lim_{N \to \infty} \tilde Q_L(-N + \delta,N) &= 
    \frac12 \begin{pmatrix}
        1 & 0 \\ 0 & 1
    \end{pmatrix} + \frac{1-\sin\theta}{2\cos\theta} \begin{pmatrix}
        1 & 0 \\ 0 & -1
    \end{pmatrix} + q^{\delta-1} \frac{\sin\theta}{2(1+\sin\theta)} \begin{pmatrix}
        1 & 1 \\ 1 & 1
    \end{pmatrix}, \\
\end{aligned}
\end{equation}
where
\begin{equation}
    \tilde Q_L(x,N) = \begin{pmatrix}
        \cos\frac\theta2 & \sin\frac\theta2 \\ -\sin\frac\theta2 & \cos\frac\theta2
    \end{pmatrix}
    Q_L(x,N)
    \begin{pmatrix}
        \cos\frac\theta2 & -\sin\frac\theta2 \\ \sin\frac\theta2 & \cos\frac\theta2
    \end{pmatrix},
\end{equation}
We see that for $k = N - \delta$ or $k=-N+\delta$, cross terms start appearing, so the probability of left or right absorption will also show dependence on the relative phase $\varphi$ in the coin eigenbasis:
\begin{equation}
\begin{aligned}
    P_L^{(\psi_c)}(\delta) & \coloneq \lim_{N\to\infty} P_L^{(\psi_c)}(N-\delta) = \frac12
      + \frac{1-\sin\theta}{2\cos\theta} (|u|^2 - |v|^2)
      + q^{\delta-1} \frac{\sin\theta}{2(1+\sin\theta)}
      (\bar u v + \bar v u - 1) \\
      & = P_L^{(\psi_c)}
      + q^{\delta-1} \frac{\sin\theta}{2(1+\sin\theta)}
      (\sin\rho \cos\varphi - 1) ,\\
   P_L^{(\psi_c)}(-\delta) & \coloneq \lim_{N\to\infty} P_L^{(\psi_c)}(-N+\delta) = \frac12
      + \frac{1-\sin\theta}{2\cos\theta} (|u|^2 - |v|^2)
      + q^{\delta-1} \frac{\sin\theta}{2(1+\sin\theta)}
      (\bar u v + \bar v u + 1) \\
      & = P_L^{(\psi_c)}
      + q^{\delta-1} \frac{\sin\theta}{2(1+\sin\theta)}
      (\sin\rho\cos\varphi + 1).
\end{aligned}
\label{eq:pl-n-delta-baze}
\end{equation}
Hence, considering finite distance $\delta$ from the left or the right absorber modifies (\ref{eq:pl}) with an exponentially decreasing term. The corresponding $P_R$ probabilities are, as elsewhere, complements to unity.

\begin{figure}
    \centering
    \includegraphics[width=0.7\linewidth]{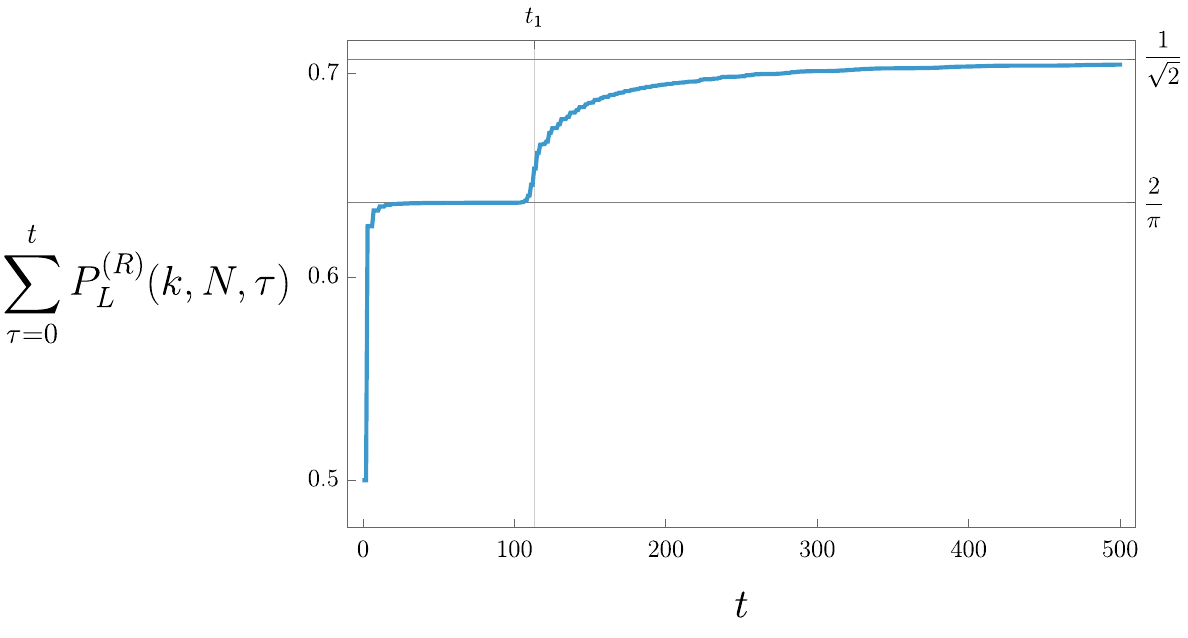}
    \caption{\newtext{Cumulative} absorption probability in the left barrier as a function of number of steps $t$ for the Hadamard walk on a line with $N=20$. The starting position was chosen next to the left absorber, i.e. $k=-19$, and the initial coin state is $\ket{R}$. We see that the absorption probability first stabilizes at a value close to $2/\pi$, corresponding to the absorption probability in the absence of the right barrier. Nevertheless, after approximately $t_1 = 4N \sqrt{2} \approx 113$ steps, the wave reflected from the right absorber reaches the left end, and the absorption probability starts to grow again, asymptotically reaching the value of $1/\sqrt{2}$.}
    \label{fig:reflection}
\end{figure}

To illustrate, we consider initial position $\delta = 1$ to the right of the left absorbing barrier as $N \to \infty$, and the initial coin state $\ket{R}$, i.e., facing away from the barrier. From \eqref{eq:ql-n-delta} we obtain
\begin{equation}
   P_L^{(R)}(-1) =  \frac12 + \frac{1-\sin\theta}{2\cos\theta} (-\cos\theta) + \frac{\sin\theta}{2(1+\sin\theta)} (1+\sin\theta) = \sin\theta .
\end{equation}
For the Hadamard coin, this is $\sin\frac{\pi}{4} = 1/\sqrt{2} \approx 0.707$, notably higher than the well-known recurrence probability $2/\pi \approx 0.637$ \cite{Ambainis2001,Grunbaum2013,Nitsche2018,sabri_2018}, which can also be reformulated as the probability of absorption in position $-1$ relative to the initial state, with the same initial coin state, except for the absence of the right barrier. We can see, even as the right barrier is moved beyond all bounds away from the initial position, its presence still reflects in $P_L$ through reflections when the propagating wave front encounters it. Direct simulation confirms that for large values of $N$, the cumulative absorption probability in \newtext{position} $-N$, \newtext{$\sum_{\tau = 0}^t P_L^{(R)}(k,N,\tau)$ (as a function of $t$)} first stabilizes near $2/\pi$, and only starts growing towards $1/\sqrt{2}$ after sufficient time has passed for the walk to cover the distance $2N$ and back, see Figure~\ref{fig:reflection} for illustration. (We note that this difference between a very distant absorber versus none at all was observed and discussed in \cite{Ambainis2001} after Theorem 10.) The time $t_1$ required for the wave traveling initially to the right to bounce from the absorber at $N$ and return back to the left absorber can be estimated by $t_1 = 4 N/v$, where $v$ is the group velocity of the peak \cite{kempf:2009}.  
For quantum walks with the coin (\ref{coin}) we have $v= \cos\theta$, reducing to $1/\sqrt{2}$ for the Hadamard walk. This shows that the order of considering the limits $t\to\infty$ and $N\to\infty$ matters.

Another interesting example is also with $\delta = 1$  to the right of the left absorbing barrier, but this time we choose the initial coin state
\begin{equation}
    \ket{\phi_L} = \hat C(\theta)^{-1} \ket{L} = \cos\theta \ket{L} + \sin\theta \ket{R}.
\end{equation}
Using \eqref{eq:ql-n-delta} we find
\begin{equation}
   P_L^{(\phi_L)} (-1)
    = \begin{pmatrix}
        \cos\theta & \sin\theta
    \end{pmatrix} \begin{pmatrix}
        \frac{1+\sin\theta-\sin^2\theta}{1+\sin\theta} & \frac{\cos\theta\sin\theta}{1+\sin\theta} \\
        \frac{\cos\theta\sin\theta}{1+\sin\theta} & \sin\theta
    \end{pmatrix} \begin{pmatrix}
        \cos\theta \\ \sin\theta
    \end{pmatrix} = 1,
\end{equation}
confirming the fact that the first application of $\hat U$ maps this initial state to $\ket{{-N},L}$ and the walker is completely absorbed in step one.

\section{Small system behavior and applicability of the analytical results}
\label{sec:numerical}

The derivation of formulas (\ref{eq:pl}) and (\ref{eq:pl-n-delta-baze}) utilized the limit of large system size $N$. For finite $N$ and $k$ we propose the following hypothesis: the \newtext{exact left absorption probability $P_L^{(\psi_c)}(k,N$)} can be well approximated by a combination of (\ref{eq:pl}) and (\ref{eq:pl-n-delta-baze}), where the initial position $k$ is considered at distance $\delta = N-k$ from the right absorber, and $\delta = N+k$ from the left absorber, namely
\newtext{
\begin{eqnarray}
\label{pl:k:approx}   \widetilde{P}_L^{(\psi_c)}(k,N) & = & P_L^{(\psi_c)}(N-k) + P_L^{(\psi_c)}(-N-k) - P_L^{(\psi_c)} \\
\nonumber & = & \frac12
      + \frac{1-\sin\theta}{2\cos\theta} \cos\rho
      + q^{N-1} \frac{\sin\theta}{2(1+\sin\theta)} \left((q^{k}+q^{-k}) 
      \sin\rho \cos\varphi + q^{k}-q^{-k}\right).
\end{eqnarray}}

\newtext{It is worth a remark that the chosen physical system, along with modeling the particle loss with a projection onto the surviving subspace (see \eqref{eq:proj-dynamics}), admits a very efficient numerical simulation. Being a single-particle system, with a finite-dimensional state space linear in $N$, and remaining in a pure state (only sub-normalized) at all times, simulating the system on a classical computer allows simulations of many thousands of time steps without significant computational cost or numerical error accumulation. Moreover, the survival probability \eqref{eq:proj-dynamics} provides a direct indicator of convergence to the limit values. In practice, the subsequent results presented as asymptotic were obtained by simulating the system universally to $T = 2 N \cdot 10^3 $ steps, after which the remaining survival probability was below double-precision machine epsilon in all cases considered below, meaning that $\sum_{t=0}^T P_L(k,N,t) + \sum_{t=0}^T P_R(k,N,t) \doteq 1$ to all significant digits, and neither of the quantities can increase any further.}

\newtext{First, we focus on the symmetric initial position, i.e. $k=0$, where the approximation (\ref{pl:k:approx}) simplifies into
\begin{eqnarray}
\label{pl:0:approx}  \widetilde{P}_L^{(\psi_c)}(0,N)  & = & \frac12
      + \frac{1-\sin\theta}{2\cos\theta} \cos\rho
      + q^{N-1} \frac{\sin\theta}{1+\sin\theta} 
      \sin\rho \cos\varphi .
\end{eqnarray}}

\begin{figure}
    \centering
    \includegraphics[width=0.6\linewidth]{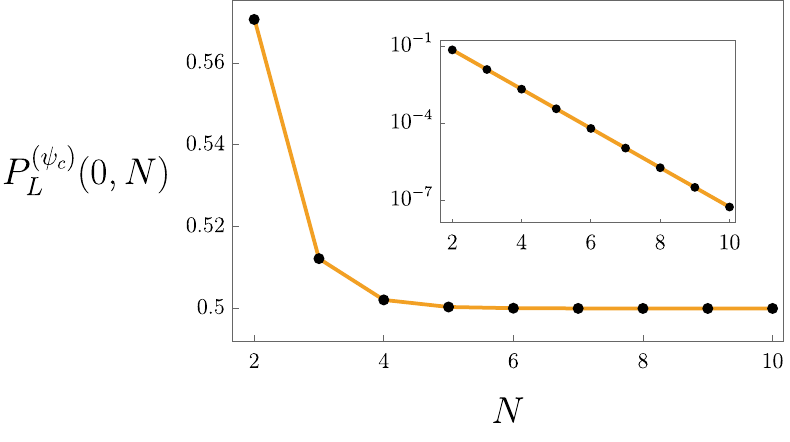}
    \caption{Numerically calculated absorption probability $P_L^{(\psi_c)}(0,N)$ for the Hadamard walk ($\theta=\frac{\pi}{4}$) starting in the middle ($k=0$) of the line as a function of $N$. The initial state is given by (\ref{init:fig3}). The inset shows the \newtext{absolute} difference from the asymptotic value \newtext{$P_L^{(\psi_c)}=\frac{1}{2}$}  on a logarithmic scale. \newtext{In both cases, the black dots show the result of numerical simulation and the orange lines the prediction obtained from \eqref{pl:0:approx}.}}
    \label{fig:conv_had}
\end{figure}

\newtext{As a first test, we study the convergence of the absorption probability $P_L^{(\psi_c)}(0,N)$ to the analytical result for constant $k$ (\ref{eq:pl}) with increasing $N$. In Figure~\ref{fig:conv_had} we consider the Hadamard walk ($\theta=\frac{\pi}{4}$) and compare the actual behaviour of $P_L(k,N)$ to that predicted by \eqref{pl:0:approx}.}
The initial coin state is chosen with angles $\rho=\pi/2$ and $\varphi = 0$,
\begin{equation}
\label{init:fig3}
    \ket{\psi_c} = \frac{1}{\sqrt{2}}\left(\ket{\theta^+} + \ket{\theta^-}\right) = \frac{\sqrt{2-\sqrt{2}}}{2} \ket{L} + \frac{\sqrt{2+\sqrt{2}}}{2} \ket{R},
\end{equation}
for which  (\ref{eq:pl}) \newtext{becomes a constant} $P_L^{(\psi_c)}=\frac{1}{2}$. The black dots are obtained from numerical simulation, the solid line is given by the formula (\ref{pl:0:approx})\newtext{, showing that the hypothesized correction terms present a viable model of the actual convergence rate.}

\begin{figure}[H]
    \centering
    \includegraphics[width=0.53\linewidth]{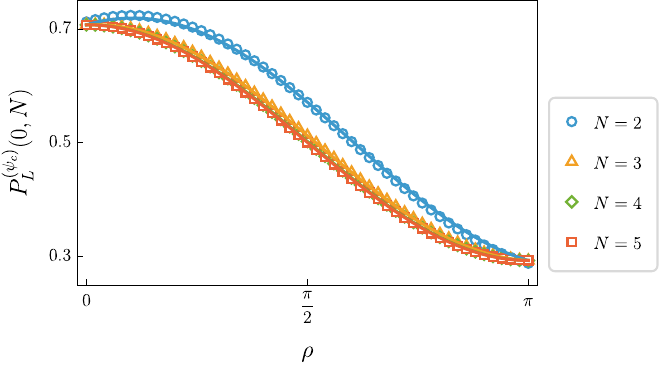}\hfill
    \includegraphics[width=0.41\linewidth]{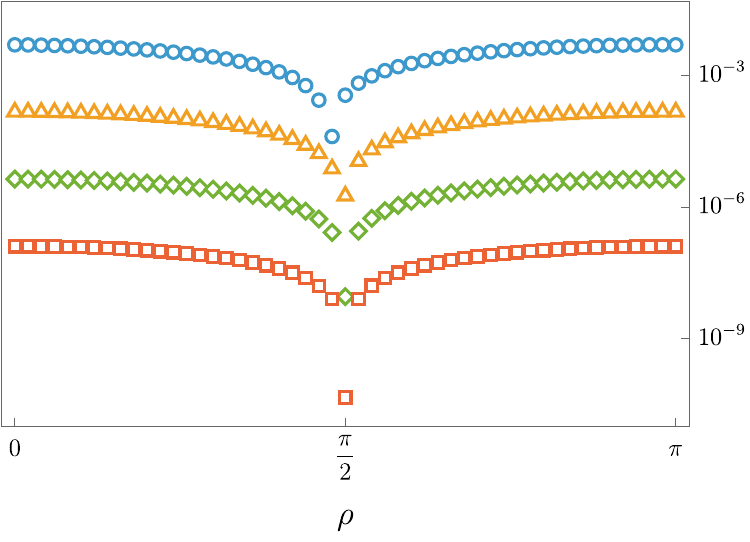}
    \caption{\newtext{On the left we show the absorption probability $P_L^{(\psi_c)}(0,N)$ for the Hadamard walk as a function of the polar angle of the initial coin state $\rho$ for $N=2,\ldots 5$. The azimuthal angle $\varphi$ is fixed to 0. The symbols are obtained from numerical simulation, the curves correspond to the approximation (\ref{pl:0:had:rho}). The right plot shows the absolute difference $\Delta$ (\ref{abs:dif:k}) on a logarithmic scale. }}
    \label{fig:had_n2}
\end{figure}

\newtext{In the following tests we quantify the quality of the approximation in more detail by looking at the deviation
\begin{equation}
  \label{abs:dif:k}  
  \Delta = \left|P_L^{(\psi_c)}(k,N) - \widetilde{P}_L^{(\psi_c)}(k,N)\right|.
\end{equation}}
In Figure~\ref{fig:had_n2} we sample the initial states of the Hadamard walk (\ref{init:eigenbasis}) by choosing $\rho = \frac{j}{50} \pi$, $j=0,\ldots 50$, while keeping the azimuthal angle fixed at $\varphi=0$. \newtext{For such $\theta$ and $\varphi$, the approximation (\ref{pl:0:approx}) yields
\begin{equation}
    \label{pl:0:had:rho}  \widetilde{P}_L^{(\psi_c)}(0,N)  = \frac12
      + \frac{\sqrt{2}-1}{2} \cos\rho
      + (\sqrt{2}+1) q^N 
      \sin\rho ,
\end{equation}
where $q = 3 - 2\sqrt{2} \doteq 0.172$. The left plot shows $P_L^{(\psi_c)}(0,N)$ for $N=2,\ldots,5$, the symbols are obtained from numerical simulations and the curves are given by (\ref{pl:0:had:rho}). The right plot shows the absolute difference $\Delta$ between the exact left absorption probability and the approximation (\ref{pl:0:had:rho}) on a logarithmic scale. We see that each increase of $N$ decreases $\Delta$ by approximately $1.5$ orders of magnitude.}

\begin{figure}[H]
    \centering
    \includegraphics[width=0.53\linewidth]{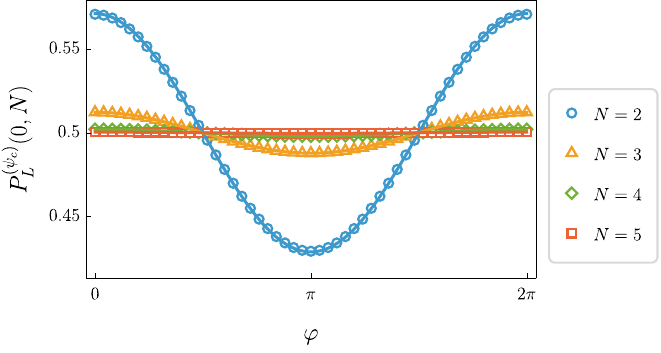}\hfill
      \includegraphics[width=0.41\linewidth]{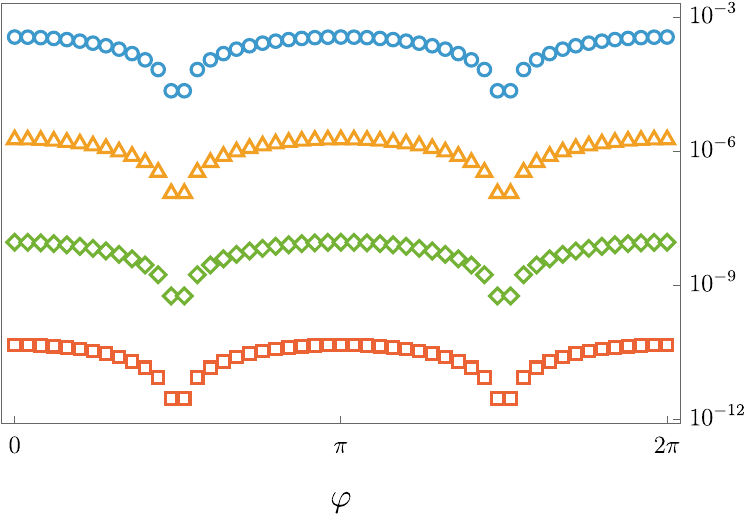}
    \caption{\newtext{On the left we plot $P_L^{(\psi_c)}(0,N)$ as a function of the relative phase \( \varphi \in [0, 2\pi] \) of the initial coin state (\ref{init:eigenbasis}) for the Hadamard walk and $N=2,\ldots, 5$. The polar angle of the initial coin state is fixed to \( \rho = \pi/2 \). Symbols correspond to the numerical simulations, curves are given by (\ref{pl:0:had:phi}). The right plot shows the absolute difference $\Delta$ (\ref{abs:dif:k}) on a logarithmic scale.}}
    \label{fig; phase role}
\end{figure}

The role of the relative phase $\varphi$ between the coin eigenstates (\ref{init:eigenbasis}) for the Hadamard walk is highlighted in Figure~\ref{fig; phase role}. \newtext{Specifically, we set $\rho = \frac{\pi}{2}$ and for $N = 2,\ldots, 5$ numerically calculate the absorption probability on the left $P_L^{(\psi_c)}(0,N)$ as a function of the relative phase $\varphi$, which is sampled as $\varphi=\frac{j}{50} 2\pi $, $j=0,\ldots,50$. For this choice of $\rho$ and $\theta$, the approximation (\ref{pl:0:approx}) reduces to
\begin{equation}
    \label{pl:0:had:phi}  \widetilde{P}_L^{(\psi_c)}(0,N)  = \frac12
      + (\sqrt{2}+1)\left(3-2\sqrt{2}\right)^N \cos\varphi.
\end{equation}
Note that the insensitivity of $P_L$ on $\varphi$, as appears in \eqref{eq:pl}, only manifests in the $N \to \infty$ limit.
In the left plot the symbols show the numerically evaluated $P_L^{(\psi_c)}(0,N)$ for $N=2,\ldots, 5$ and curves are given by (\ref{pl:0:had:phi}). The right plot shows the absolute difference $\Delta$ (\ref{abs:dif:k}) on a logarithmic scale. In this case increasing $N$ decreases $\Delta$ by more than 2 orders of magnitude, although judging by Figure~\ref{fig:had_n2}, this faster rate appears to be specific to the choice of $\rho = \frac{\pi}{2}$.}

\begin{figure}[H]
    \centering
    \includegraphics[width=0.46\linewidth]{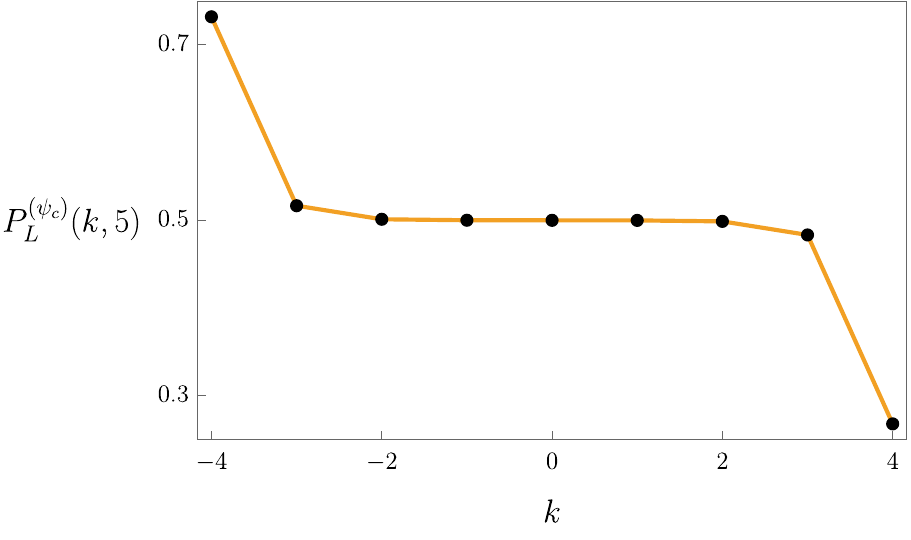}\hfill
    \includegraphics[width=0.49\linewidth]{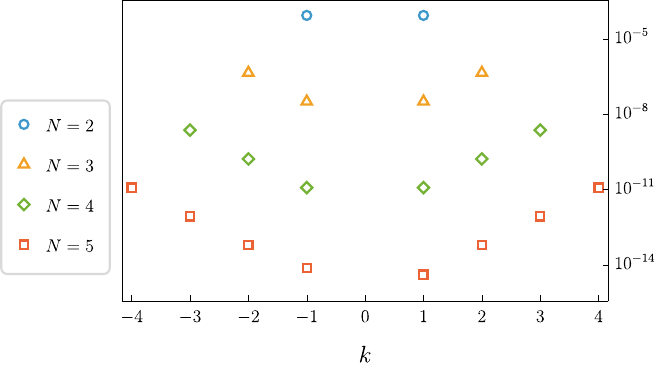}
    \caption{\newtext{Absorption probability $P_L^{(\psi_c)}(k,N)$ as a function of the initial position $k$. The initial coin state is given by (\ref{initc:fig7}) and the coin angle equals $\theta = \pi/3$. The left plot shows the results for $N = 5$, the black dots are obtained from numerical simulation, the orange curve is given by (\ref{pl:k:approx:fig7}). The right plot shows the absolute difference $\Delta$ between the exact value and (\ref{pl:k:approx:fig7}) on the logarithmic scale for $N=2,\ldots ,5$ and $k=-N+1,\ldots ,N-1$. The difference vanishes for $k = 0$.}}
    \label{fig:7}  
\end{figure}

\begin{figure}[H]
    \centering
    \includegraphics[width=0.46\linewidth]{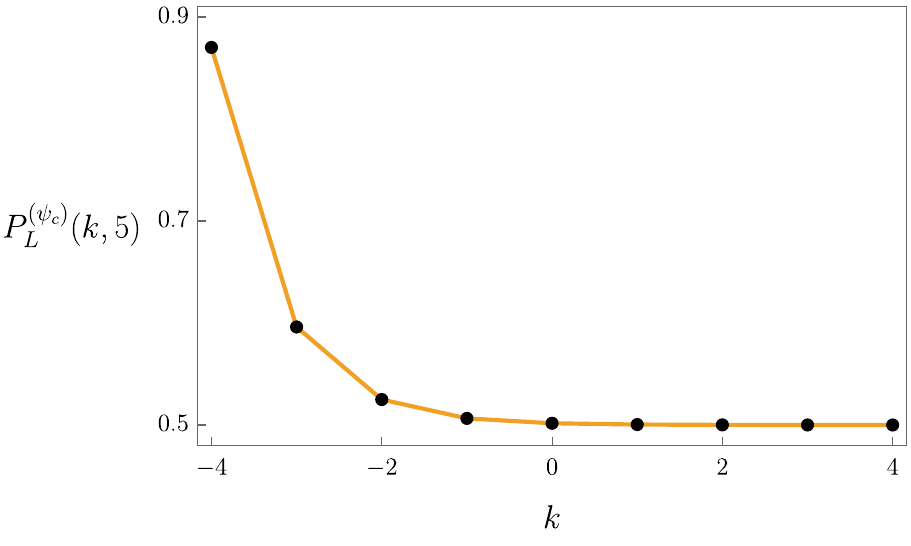}\hfill
    \includegraphics[width=0.49\linewidth]{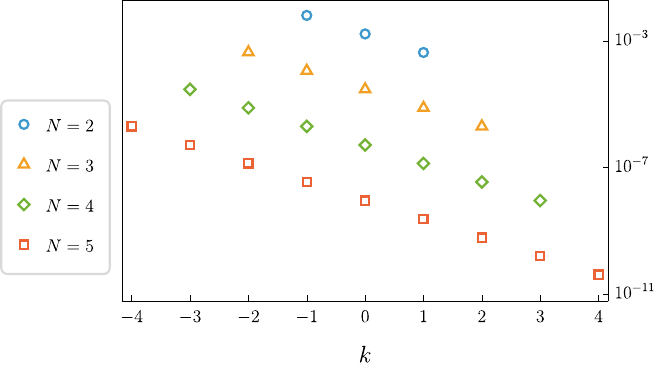}
    \caption{\newtext{On the left we display the absorption probability $P_L^{(\psi_c)}(k,N)$ as a function of the initial position $k$. We consider $N = 5$ and the coin angle $\theta = \pi/5$. The initial coin state is given by (\ref{initc:fig8}). Black dots are obtained from numerical simulation, orange curve corresponds to (\ref{pl:k:approx:fig8}). The right plot shows the absolute difference $\Delta$ between the exact absorption probability and the approximation (\ref{pl:k:approx:fig8}) on the logarithmic scale for $N=2,\ldots, 5$.}}
    \label{fig:8}  
\end{figure}

Let us now consider varying initial position $k$ which we illustrate in Figures~\ref{fig:7} and \ref{fig:8}. In Figure~\ref{fig:7} we consider the initial coin state with parameters $\rho = \pi/2$ and $\varphi = \pi/2$
\begin{equation}
\label{initc:fig7}
    \ket{\psi_c} = \frac{1}{\sqrt{2}}\left(\ket{\theta^+} + i\ket{\theta^-}\right),
\end{equation}
where (\ref{pl:k:approx}) reduces to
\newtext{
\begin{eqnarray}
\label{pl:k:approx:fig7}   \widetilde{P}_L^{(\psi_c)}(k,N) & = &  \frac12 + q^{N-1} \frac{\sin\theta}{2(1+\sin\theta)} \left(q^{k}-q^{-k}\right).
\end{eqnarray}}
\newtext{For Figure~\ref{fig:7} we have chosen the coin angle $\theta = \pi/3$, where $q = 7-4\sqrt{3} \doteq 0.0718$. In the left plot we show $P_L^{(\psi_c)}(k,N)$ for $N=5$. Black dots are obtained from numerical simulation, solid curve corresponds to (\ref{pl:k:approx:fig7}). }
We see that for negative $k$, the term $q^{-k}$ is negligible, and $q^{k}$ increases the left absorption probability above the value $1/2$. The difference drops exponentially quickly as $k$ increases towards 0. For positive $k$, the roles are interchanged, i.e. $q^{k}$ is negligible, and $q^{-k}$ decreases $P_L^{(\psi_c)}(k,N)$ below $1/2$, with the difference growing exponentially in absolute value. \newtext{The right plot shows the absolute difference $\Delta$  (\ref{abs:dif:k}) on the logarithmic scale for $N=2,\ldots, 5$.}

\newtext{The separate appearance of the terms $q^k$, $q^{-k}$, responsible for the two exponential modulations seen in the preceding discussion, in the formula \eqref{pl:k:approx} gives a simple way of identifying a parameter regime in which one of them vanishes, namely $\sin\rho \cos\varphi = \pm 1$.
In Figure~\ref{fig:8} we explore this phenomenon by considering the initial coin state with $\rho=\pi/2$, $\varphi=0$,
\begin{equation}
\label{initc:fig8}
   \ket{\psi_c} = \frac{1}{\sqrt{2}}\left(\ket{\theta^+} + \ket{\theta^-}\right).
\end{equation}
In this case the formula (\ref{pl:k:approx}) gives
\begin{eqnarray}
\label{pl:k:approx:fig8}   \widetilde{P}_L^{(\psi_c)}(k,N)  =  \frac12 + q^{N+k-1} \frac{\sin\theta}{1+\sin\theta}.
\end{eqnarray}
We see that the left absorption probability follows a simple exponential decrease with $k$, starting from $\frac12 +\frac{\sin\theta}{1+\sin\theta}$ at $k=-N+1$ and approaching close to $\frac12$ for $k>0$.
In the plot we consider the coin angle $\theta = \pi/5$, which gives $q\doteq 0.26$. Note that for $\varphi = \pi$ we obtain
\begin{eqnarray}
 \widetilde{P}_L^{(\psi_c)}(k,N)  = \frac12 - q^{N-k-1} \frac{\sin\theta}{1+\sin\theta},
\end{eqnarray}
so in this case the other of the two exponentials is suppressed: the left absorption probability starts close to $\frac12$ for $k=-N+1$ and decreases towards $\frac12 - \frac{\sin\theta}{1+\sin\theta}$ at $k=N-1$. }

\begin{figure}[H]
    \centering
    \includegraphics[width=0.56\linewidth]{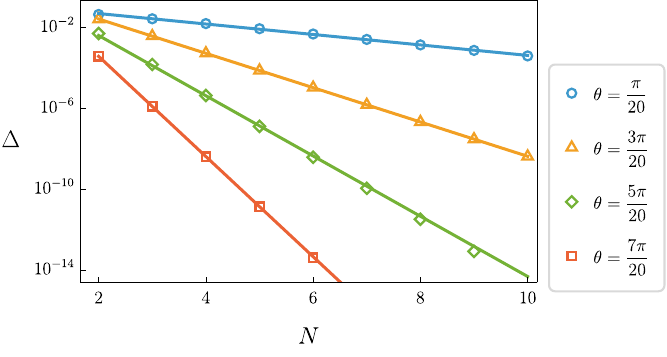}\hfill
    \includegraphics[width=0.43\linewidth]{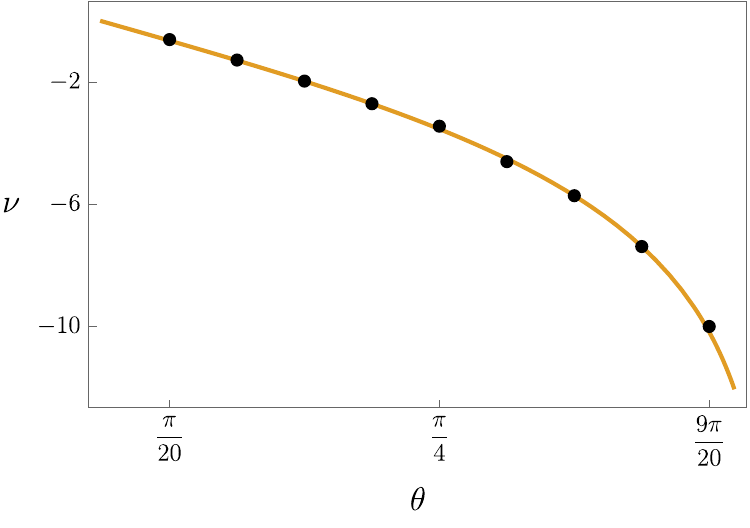}
    \caption{\newtext{On the left we quantify the decrease of the absolute difference $\Delta$ in $N$ for several different choices of the coin parameter $\theta$. The plot admits a good fit by $\text{const.} \exp(\nu N)$. The right plot shows the fitted value of this $\nu$ for a span of coin parameters covering most of the admitted interval $(0, \pi/2)$. Interestingly, the empirical data is nicely interpolated by $2 \log q$, i.e., $\exp(\nu N) \approx q^{2N}$, plotted as the orange line.}}
    \label{fig:9}  
\end{figure}

\newtext{Finally we investigate the dependence on the coin parameter $\theta$. We remind that the theory in Section~\ref{sec:analytical} was derived under the assumption $\theta \in (0, \pi/2)$, and that the boundary cases display qualitatively different behaviour, as discussed already in Section~\ref{sec:overview}. For this reason we restrict our analysis to an interior domain to keep the numerical evaluation away from the endpoints. Figure~\ref{fig:9} shows the difference $\Delta$ in several combinations of $\theta$ and $N$ for the initial state $\ket{\theta^+}$ ($\rho = \varphi = 0$), which was observed to represent the highest deviation for initial position $k = 0$ in Figure~\ref{fig:had_n2}. A clear exponential decrease in $N$ can be identified, with its quotient depending on $\theta$. The agreement with the function \eqref{pl:k:approx} is seen to be better, both in small line sizes and in the decrease of the deviation as $N$ grows, as $\theta$ gets closer to the right endpoint, i.e., for walks with slower spreading.}

\section{Conclusions}
\label{sec:conclusions}

Absorption of two-state quantum walks on a finite line was investigated in detail. Utilizing the two-scale convergence analysis, we derived closed formulas for the absorption probabilities in the limit of large system size $N$. The results have particularly simple form when the initial coin state is expressed in the eigenbasis of the coin operator. In the case when the starting position $k$ is fixed during the limit process, the absorption probability is independent of the relative phase in the initial state. Considering the initial vertex at distance $\delta$ from the left or right barrier contributes with a correction exponentially decreasing with $\delta$. Inspired by the analytical results we proposed an approximation (\ref{pl:k:approx}) for the absorption probability for finite $k$ and $N$. We found that the approximation is in excellent agreement with the numerical simulations even for small $N$. \newtext{The absolute difference between the exact absorption probability and the approximation (\ref{pl:k:approx}) further decreases exponentially with $N$. %The decay rate depends on the coin angle $\theta$ and can be well approximated by $\nu \approx 2\log q$.
We note that while we have considered lines of odd lengths, all derived results can be easily adopted to lines of even length as well. The details are left for Appendix \ref{app:c}.}

As a possible generalization, one can consider lazy walk model on a line which leads to trapping \cite{Inui2005}, where the walker has a non-vanishing probability to survive even in the asymptotic limit of large number of steps. While the total absorption probability was investigated before \cite{stefanak:2016}, the absorption in the left and right sink individually is not known. We can also investigate absorption problems with multiple sinks in graphs different from a path. The generating function approach utilized for the study of recurrence \cite{Grunbaum2013,gruenbaum2014,grunbaum:2018,grunbaum:recurrence:2020,stefanak:rec:2025} can be extended to investigate absorption in individual sinks.

Finally, it would be interesting to see an experimental realization of quantum walks models with absorption. The photonic time-multiplexing setup \cite{Nitsche2016} provides a variable platform which allows to investigate different system sizes $N$ in a straightforward way. Sinks at desired positions $\pm N$ can be implemented with deterministic out-coupling \cite{Nitsche2018} utilizing programmable electro-optical modulators.

\begin{acknowledgments}
A. A. and F. V. P. acknowledge support from Ministero dell’Universit\`{a} e della Ricerca (MUR) PNRR project CN00000013 ``National Centre on HPC, Big Data and Quantum Computing''. F. V. P. acknowledges support from Ministero dell’Universit\`{a} e della Ricerca (MUR) PNRR project PE0000023259 ``National Quantum Science and Technology Institute (NQSTI)''. V. P. and M. \v{S}. have been supported by the Grant Agency of the Czech Republic GAČR under Grant No. 23-07169S.
\end{acknowledgments}

\appendix

\section*{Appendix}

\section{Simplification of integrands}
\label{sec:integrands}

In this Appendix we simplify the integrands of coefficients $C_j(k,N)$ (\ref{eq:cj:coef}). First, note that for $z = e^{i \phi}$, $\lambda_\pm$ \eqref{eq:lambda} can be written in the form
\begin{equation}
        \lambda_\pm = i \frac{\sin\phi}{\cos\theta} \pm \sqrt{1 - \left(\frac{\sin\phi}{\cos\theta}\right)^2} = i \cos\alpha \pm \sin\alpha = i e^{\pm i \alpha},
\end{equation}
where
\begin{equation}
    \alpha = \arccos \frac{\sin\phi}{\cos\theta}.
    \label{eq:alpha}
\end{equation}
($\alpha$ taken to be $i \mu$ or $\pi + i\mu$, $\mu \in \mathbb{R}_+$, if $\sin\phi > \cos\theta$ or $\sin \phi < -\cos\theta$, respectively.)

The constructions \eqref{eq:x,y} then become
\begin{equation}
    x_n = 2 i^n \cos(n \alpha), \quad y_n = 2 i^{n+1} \sin(n \alpha),
\end{equation}
valid also in the cases of $\alpha \not \in \mathbb{R}$, with the goniometric functions becoming hyperbolic in $\mu$.

In these terms, the solution of (\ref{eq:cz:ez}) can be found as
\begin{equation}
\begin{aligned}
    C_z &= \frac{(-1)^N z^2 \sin\theta}{2\left(\sin((2N-1)\alpha) + i z \cos\theta \sin((2N-2)\alpha) \right)} , \\ 
    E_z &= \frac{z}{2}\frac{i \cos((2N-1)\alpha) - z \cos\theta \cos((2N-2)\alpha)}{\sin((2N-1)\alpha) + i z \cos\theta \sin((2N-2)\alpha)}.
\end{aligned}
\end{equation}
Using this result in equations (\ref{eq:pn:rn}) we find that the functions $p^{(2N)}_{N+k}(z)$ and $r^{(2N)}_{N+k}(z)$ are given by
\begin{equation}
\begin{aligned}
\label{sol:pn:rn:odd}
    p^{(2N)}_{N+k}(z) &= i^{N+k} z \frac{z \cos\theta \sin((N-k-1)\alpha) - i \sin((N-k)\alpha)}{\sin((2N-1)\alpha) + i z \cos\theta \sin((2N-2)\alpha)}, \\ 
    r^{(2N)}_{N+k}(z) &= i^{N+k} z^2 \frac{\sin\theta \sin((N-k-1)\alpha)}{\sin((2N-1)\alpha) + i z \cos\theta \sin((2N-2)\alpha)}.
\end{aligned}
\end{equation}

Turning finally to the coefficients (\ref{eq:cj:coef}), we express the integrand of $C_1(k,N)$ as
\begin{equation}
    I_1(\phi;k,N) = \frac{|\corr{e^{i \phi}} \sin((N-k-1)\alpha) - i \cos\theta \sin((N-k)\alpha)|^2}{|\sin((2N-1)\alpha) + i \corr{e^{i \phi}} \cos\theta \sin((2N-2)\alpha)|^2}
\end{equation}
Given that the sine terms are either all real or all pure imaginary, it's easy to separate the real and imaginary parts and arrive at
\begin{equation}
\begin{aligned}
    I_1(\phi;k,N) &= \frac{\sin((N-k-1)\alpha)^2 + \cos^2\theta \sin((N-k)\alpha)^2 - 2\sin\phi \cos\theta \sin((N-k)\alpha) \sin((N-k-1)\alpha)}{\sin((2N-1)\alpha)^2 + \cos^2\theta \sin((2N-2)\alpha)^2 - 2\sin\phi\cos\theta\sin((2N-1)\alpha) \sin((2N-2)\alpha)} \\
    &= \frac{1 - \cos^2\theta \cos(2\alpha) - \sin^2\theta \cos((2N-2k-2)\alpha)}{1 - \cos^2\theta \cos(2\alpha) - \sin^2\theta \cos((4N-2)\alpha)}
\end{aligned}
\end{equation}
using the property \eqref{eq:alpha} that $\sin\phi = \cos\theta\cos\alpha$ and expanding all powers and products of the $\alpha$-dependent terms.

Analogously we express the integrands of $C_2(k,N)$ and $C_3(k,N)$ as
\begin{equation}
\begin{aligned}
    I_2(\phi;k,N) &= \frac{\sin^2\theta (1 - \cos((2N-2k)\alpha))}{1 - \cos^2\theta \cos(2\alpha) - \sin^2\theta \cos((4N-2)\alpha)}, \\
    I_3(\phi;k,N) &= \frac{\frac12\sin\theta\cos\theta (1 - \cos(2\alpha) - \cos((2N-2k)\alpha) + \cos((2N-2k-2)\alpha))}{1 - \cos^2\theta \cos(2\alpha) - \sin^2\theta \cos((4N-2)\alpha)} +\\
    &\quad + \frac{2i \cos\phi \sin\theta (\cos\alpha - \cos((2N-2k-1)\alpha))}{1 - \cos^2\theta \cos(2\alpha) - \sin^2\theta \cos((4N-2)\alpha)}
\end{aligned}
\end{equation}
Here, we separated the real and imaginary parts of $I_3$. Since it can be seen that the latter is an odd function in $\phi$ (through the odd-multiple cosines of \eqref{eq:alpha}), its integral over $(0,2\pi)$ yields zero. Hence, $C_3(k,N)$ is real valued, and it is determined by integrating the real part of $I_3$ only.

\section{Evaluation of integral $A_m$}
\label{app:b}

In this Appendix we show how the integral (\ref{eq:Am})
\begin{equation}
    A_m = \frac{1}{2\pi} \int_0^{2\pi} \frac{\cos(2m\alpha)}{2 (1-\cos^2\theta\cos^2\alpha)} d\alpha , 
\end{equation}
can be calculated for $m \in \mathbb{N}_0$. We re-express using $\zeta = e^{i \alpha}$ as
\begin{equation}
\begin{aligned}
    A_m &= \frac{1}{2\pi} \Re \int_0^{2\pi} \frac{\zeta^{2m}}{2 \left(1-\cos^2\theta\left(\frac{\zeta+\zeta^{-1}}{2}\right)^2\right)} d\alpha = \frac{1}{2\pi} \Re \int_0^{2\pi} \frac{2\zeta^{2m}}{4-\cos^2\theta\left(\zeta^2+\zeta^{-2}+2\right)} d\alpha \\
    &= \frac{1}{2\pi} \Im \int_0^{2\pi} \frac{2\zeta^{2m+1}}{4\zeta^2-\cos^2\theta\left(\zeta^4+2\zeta^2+1\right)} i \zeta d\alpha \\
\end{aligned}
\end{equation}
and ultimately as a contour integral
\begin{equation}
    A_m = \frac{1}{2\pi} \Im \oint \frac{2\zeta^{2m+1}}{4\zeta^2-\cos^2\theta\left(\zeta^4+2\zeta^2+1\right)} d\zeta
\end{equation}
over the unit circle. The denominator has four simple zeroes as given by
\begin{equation}
    \zeta^2 = \frac{2-\cos^2\theta \pm \sqrt{4-4\cos^2\theta}}{\cos^2\theta} = \frac{(1\pm\sin\theta)^2}{\cos^2\theta}
\end{equation}
out of which only
\begin{eqnarray}
    \zeta_\pm = \pm \frac{1-\sin\theta}{\cos\theta}
\end{eqnarray}
lie within the unit disk. By the residue theorem, $A_m$ is then equal to
\begin{equation}
    A_m = \sum_{\zeta=\zeta_\pm} 2\zeta^{2m+1} \mathop{\mathrm{Res}}\nolimits_\zeta \frac{1}{4\zeta^2-\cos^2\theta\left(\zeta^4+1+2\zeta^2\right)}
\end{equation}
Having evaluated all roots of the denominator, the residue is easy to calculate:
\begin{eqnarray}
\nonumber  \mathop{\mathrm{Res}}\nolimits_{\zeta_\pm} \frac{1}{4\zeta^2-\cos^2\theta\left(\zeta^4+1+2\zeta^2\right)} & = & -\frac{1}{\cos^2\theta} \frac1{\zeta_\pm - \zeta_\mp} \frac1{\zeta_\pm - \frac{1+\sin\theta}{\cos\theta}}\frac1{\zeta_\pm + \frac{1+\sin\theta}{\cos\theta}}
 \\
    & = & \pm\frac{\cos\theta}{8\sin\theta(1-\sin\theta)},
\end{eqnarray}
and hence
\begin{eqnarray}
\nonumber    A_m & = & 2\zeta_+^{2m+1} \frac{\cos\theta}{8\sin\theta(1-\sin\theta)} - 2\zeta_-^{2m+1} \frac{\cos\theta}{8\sin\theta(1-\sin\theta)} = \frac{1}{2\sin\theta} \left(\frac{1-\sin\theta}{\cos\theta}\right)^{2m}\\
    & = & \frac{1}{2\sin\theta} \left(\frac{1-\sin\theta}{1+\sin\theta}\right)^m.
\end{eqnarray}

\newtext{\section{Absorption on a line of even length}
\label{app:c}
The derivation in the main text considered the situation of absorbers located in $-N$ and $N$, which, by a change of coordinates, can describe any line segment of odd length (in terms of number of vertices). However, there is no fundamental limitation to the validity of the obtained results to lines of even length, except that a minor point regarding the parity symmetry gets more complicated.

For a line of even length, we can similarly assume, without loss of generality, the absorbers at vertices $-N$ and $N+1$. According to \cite{Konno:absorp:2003} the equations (\ref{eq:pn:rn}) and (\ref{eq:cz:ez}) have to be modified as follows
\begin{align}
    \nonumber p^{(2N+1)}_{N+k}(z) & = \frac{z}{2} x_{N+k-1} + E_z y_{N+k-1}, \\
\label{eq:pn:rn2}    r^{(2N+1)}_{N+k}(z) & = (-1)^{N-k+1} C_z y_{N-k},
\end{align}
and
\begin{align}
    \nonumber C_z\, y_{2N-1} & = z \cos\theta\, C_z\, y_{2N-2} - z \sin\theta\, \left(\frac{z}{2}\, x_1 + E_z\, y_1\right) , \\
\label{eq:cz:ez2}    C_z\, y_1 & = z \sin\theta\, \left(\frac{z}{2}\, x_{2N-1} + E_z\, y_{2N-1}\right).
\end{align}
Solving equations (\ref{eq:cz:ez2}) and utilizing the substitution (\ref{eq:alpha}) we obtain
\begin{equation}
\begin{aligned}
    p^{(2N+1)}_{N+k}(z) &= i^{N+k} z \frac{z \cos\theta \sin((N-k)\alpha) - i \sin((N-k+1)\alpha)}{\sin(2N\alpha) + i z \cos\theta \sin((2N-1)\alpha)}, \\ 
    r^{(2N+1)}_{N+k}(z) &= i^{N+k} z^2 \frac{\sin\theta \sin((N-k)\alpha)}{\sin(2N\alpha) + i z \cos\theta \sin((2N-1)\alpha)}.
\end{aligned}
\end{equation}
In comparison with the result for odd length of the line (\ref{sol:pn:rn:odd}) we see that there is one more $\alpha$ in all arguments of sines. A similar simple change then applies to the integrands $I_j(\phi;k,N)$:
\begin{equation}
\begin{aligned}
    I_1(\phi;k,N/N+1) &=  \frac{1 - \cos^2\theta \cos(2\alpha) - \sin^2\theta \cos((2N-2k)\alpha)}{1 - \cos^2\theta \cos(2\alpha) - \sin^2\theta \cos(4N\alpha)}, \\
     I_2(\phi;k,N/N+1) &= \frac{\sin^2\theta (1 - \cos((2N-2k+2)\alpha))}{1 - \cos^2\theta \cos(2\alpha) - \sin^2\theta \cos(4N\alpha)}, \\
    \mathrm{Re}\ I_3(\phi;k,N/N+1) &= \frac{\frac12\sin\theta\cos\theta (1 - \cos(2\alpha) - \cos((2N-2k+2)\alpha) + \cos((2N-2k)\alpha))}{1 - \cos^2\theta \cos(2\alpha) - \sin^2\theta \cos(4N\alpha)} .
    \label{eq:i1even}
\end{aligned}
\end{equation}
Since this change is strictly localized to the $N$-dependent terms in $\alpha$, the complete analogy of the limit procedure done in Section~\ref{sec:analytical} holds in the case $k \ge 0$. The result \eqref{eq:result-const} for $k = \mathrm{const.}$ holds without any change, and in \eqref{eq:result-n-delta} the meaning of $\delta$ only needs to be updated to $\delta = (N+1) - k$, the distance from the right absorber.

The only technical complication happens in the regime $k < 0$, which in the case of odd length was delegated to the solved case with the help of the parity symmetry. For a line with absorbers at $-N$ and $N+1$, the mapping prescribed by \eqref{eq:parity-op} is not a symmetry of the system, because the boundary conditions are not symmetric around zero position. Nevertheless, the parity operator may still be applied, mapping between the system in question and another instance of the problem which has absorbers at $N$ and $-(N+1)$. Modifying \eqref{eq:parity-op} slightly to
\begin{equation}
    \hat P': \ket{x} \otimes \ket{c} \mapsto \ket{{-x+1}} \otimes (\sigma_y \ket{c})
\end{equation}
shifts these back to $-N$ and $N+1$ and, indeed, becomes a new reflection symmetry for the even-length line. As it maps initial positions $k < 0$ to $k > 1$, all cases of $k$ are solved.

Finally, the expression for the hypothesized small-system approximation also only changes in the term expressing the distance from the right absorber,
\begin{eqnarray}
\label{pl:k:approx2}
\widetilde{P}_L^{(\psi_c)}(k,N) & = & P_L^{(\psi_c)}(N+1-k) + P_L^{(\psi_c)}(-N-k) - P_L^{(\psi_c)} \\
\nonumber & = & \frac12
      + \frac{1-\sin\theta}{2\cos\theta} \cos\rho
      + q^{N} \frac{\sin\theta}{2(1+\sin\theta)} \left((q^{k-1}+q^{-k}) 
      \sin\rho \cos\varphi + q^{k-1}-q^{-k}\right).
\end{eqnarray}
}

\bibliography{references}
\bibliographystyle{unsrt}

\end{document}